\documentclass[journal]{IEEEtran}

\usepackage{graphicx}
\usepackage{epstopdf}
\usepackage{amsmath}
\usepackage{color}

\begin{document}

\title{Joint Power Splitting and Relay Selection in Energy-Harvesting Communications for IoT Networks}

\markboth{IEEE Internet of Things Journal (ACCEPTED TO APPEAR)}%
{Yulong~Zou \emph{et al.}: Security-Reliability Tradeoff for Distributed Antenna Systems in Heterogeneous Cellular Networks}

\author{Yulong~Zou,~\IEEEmembership{Senior Member,~IEEE}, Jia~Zhu, and Xiao~Jiang

\thanks{Copyright (c) 2019 IEEE. Personal use of this material is permitted. However, permission to use this material for any other purposes must be obtained from the IEEE by sending a request to pubs-permissions@ieee.org.}

\thanks{Manuscript received May 15, 2019; revised August 6, 2019 and September 19, 2019; accepted October 7, 2019. The editor handling the review of this paper was Dr. Cedomir Stefanovic.}

\thanks{This work was partially supported by the National Natural Science Foundation of China (Grant Nos. 61522109, 61631020 and 91738201) and the Natural Science Foundation of Jiangsu Province (Grant Nos. BK20150040 and BK20171446).}

\thanks{Y. Zou (corresponding author), J. Zhu, and X. Jiang are with the School of Telecommunications and Information Engineering, Nanjing University of Posts and Telecommunications, Nanjing 210003, China.}

}

\maketitle

\begin{abstract}
In this paper, we consider an energy-harvesting (EH) relay system consisting of a source, a destination, and multiple EH decode-and-forward (DF) relays that can harvest the energy from their received radio signals. A power-splitting ratio is employed at an EH DF relay to characterize a trade-off between the energy used for decoding the source signal received at the relay and the remaining energy harvested for retransmitting the decode outcome. We propose an optimal power splitting based relay selection (OPS-RS) framework and also consider the conventional equal power splitting based relay selection (EPS-RS) for comparison purposes. We derive closed-form expressions of outage probability for both the OPS-RS and EPS-RS schemes and characterize their diversity gains through an asymptotic outage analysis in the high signal-to-noise ratio region. We further examine an extension of our OPS-RS framework to an energy-harvesting battery (EHB) enabled cooperative relay scenario, where the EH relays are equipped with batteries used to store their harvested energies. We propose an amplify-and-forward (AF) based EHB-OPS-RS and a DF based EHB-OPS-RS schemes for AF and DF relay networks, respectively. Numerical results show that the proposed OPS-RS always outperforms the EPS-RS in terms of outage probability. Moreover, the outage probabilities of AF and DF based EHB-OPS-RS schemes are much smaller than that of OPS-RS and EPS-RS methods, demonstrating the benefit of exploiting the batteries in EH relays. Additionally, the DF based EHB-OPS-RS substantially outperforms the AF based EHB-OPS-RS and such an outage advantage becomes more significant, as the number of EH relays increases.

\end{abstract}


\IEEEpeerreviewmaketitle

\section{Introduction}
\IEEEPARstart{W}{ireless} cooperative communications has attracted tremendous research attention in recent years [1]-[5], where a user terminal first transmits a source signal to a cooperative partner which then relays its received signal to the desired destination. As shown in [6] and [7], there are two basic relaying protocols, namely the amplify-and-forward (AF) and decode-and-forward (DF). Generally, AF allows the relay to simply retransmit a scaled version of its received noisy signal without any sort of decoding, which has an advantage of simple implementation, but suffers from a noise propagation issue. By contrast, in the DF protocol, the relay needs to decode the noisy source signal first and retransmits its correctly decoded outcome to the destination, which can alleviate the noise propagation problem at the cost of computational complexity and latency. It is generally recognized that cooperative relay communications relying on either AF or DF is capable of improving the network coverage and transmission reliability in wireless fading environments [8].

Considering multiple relays available in wireless networks, one may exploit all of them to simultaneously forward their received source signals over a single channel with the aid of beamforming, also termed as cooperative relay beamforming {{[9]}}-[11]. Although such cooperative beamforming technique can significantly improve the wireless throughput, it requires complex symbol-level relay synchronization for the sake of mitigating inter-symbol interference among spatially distributed relays. Alternatively, multiple relays may be considered to retransmit their received source signals over orthogonal channels to avoid the complex relay synchronization [12]-[14], which however sacrifices multiplexing gain and reduces spectral utilization, since more orthogonal channel resources are consumed with an increasing number of relays. To this end, opportunistic relay selection as a promising means of efficiently utilizing multiple relays has been studied extensively in literature [15]-[17], where the single best relay is chosen to retransmit the source signal and thus only one orthogonal channel is required for the best-relay retransmission regardless of the number of relays. It has been shown in [15]-[17] that the best-relay selection approach can achieve the same full diversity as the aforementioned cooperative relay methods [10]-[14].

Typically, wireless terminals are powered by rechargeable batteries with limited energy capacity. As an alternative, energy harvesting (EH) is emerging as an effective means of enabling wireless devices to capture energy from their surrounding environments, such as the solar energy, wind energy, and radio frequency (RF) energy, which is attractive to machine-to-machine (M2M) communications [18]-[20] in IoT networks [21]-[25], {{since IoT sensors and devices are generally powered by energy-limited batteries without constant power supply}}. In [26], simultaneous wireless information and power transfer (SWIPT) was investigated and two SWIPT protocols were proposed, namely the time-switching (TS) and power-splitting (PS) protocols. To be specific, the PS protocol considers the use of a power splitter to separate a received RF signal into two parts for the energy harvesting and signal decoding, respectively. By contrast, in the TS protocol, the received RF signal is divided into two components in time domain [27]. Moreover, the authors of [28] studied the impact of a nonlinear energy harvesting model on the achievable rate region of RF-powered two-way communication systems.

\subsection{Related Works and Motivation}
Recently, an increasing research attention has been paid to the combination of EH with cooperative relays [29]-[35]. More specifically, in [29], a tradeoff between the energy relaying (ER) and information relaying (IR) was investigated for the sake of maximizing the delay-limited throughput of RF powered wireless networks. The authors of [30] studied wireless transmissions with the help of a single EH relay and derived a closed-form outage probability expression for the EH relay aided cooperative transmissions over Rayleigh fading channels. Later on, in [31], an interior point penalty function based EH relay selection algorithm was presented for a multiple EH relay network. In [32], the authors derived closed-form outage probability expressions of relay selection for cooperative EH DF relay networks with a fixed power splitting factor. Differing from the DF protocol used in [32], the authors of [33] and [34] considered the AF strategy for EH relays and proposed various EH relay selection schemes to improve the outage performance of EH AF relay networks. Additionally, an extension of the EH relay selection to multi-antenna multi-relay systems was investigated in [35], where a joint relay-and-antenna selection scheme is proposed to enhance the transmission throughput and reduce the outage probability.

In addition to the aforementioned EH relay selection work [29]-[35], there are also some research efforts devoted to exploring power allocation for energy-harvesting relay systems [36]-[39]. For example, in [36], the authors studied the power allocation for a cooperative DF relay network in which the source node is assumed to have a limited energy storage and can harvest power from its surrounding RF signals. The authors of [37] considered a cooperative EH relay network consisting of multiple source-destination pairs with the assistance of an EH relay and an auction based power allocation scheme was proposed to address the harvested energy distribution among the multiple source nodes. Moreover, in [38], dynamic programming was exploited for the optimal power allocation of wireless EH communications in terms of minimizing the outage probability. Besides, the authors of [39] considered an EH relay network consisting of a source, an EH relay and a destination, where the source transmits its confidential information to the relay, while the destination sends an interference signal to prevent the relay from decoding the source message. The optimal power allocation between the source and destination was investigated to maximize the secrecy rate of source-destination transmissions.

It needs to be pointed out that the aforementioned efforts [29]-[39] have separately investigated either the relay selection or power allocation for EH relay systems. Presently, only a few research attention has been paid to the joint relay selection and resource allocation for EH cooperative networks [40], [41]. Specifically, in [40], the authors formulated a sum-rate maximization problem under the constraints of total transmit power and harvested energy for a two-way AF relay network, for which an optimal resource allocation and relay selection scheme is proposed. Moreover, the geometric programming and binary particle swarm optimization were exploited in [41] to address the joint optimization of relay selection and power splitting for an EH-based two-way AF relaying system. Different from the AF protocol considered in [40] and [41], we explore the joint power splitting and relay selection for a cooperative EH DF relay network consisting of a source, a destination, and multiple DF relays that are capable of harvesting energies from their surrounding RF environments. A power splitter is assumed at each EH relay to divide its received RF signal into two separate parts for the information decoding and relaying, respectively, where a power-splitting ratio (PSR) is defined as the ratio of an harvested energy for the information relaying to the total RF energy received at the relay.

The main contributions of this paper can be summarized as follows. First, we propose an optimal power splitting based relay selection (OPS-RS) framework for cooperative EH DF relay systems{{, which is different from [40] and [41], where the AF protocol was considered for the joint power splitting and relay selection design. This also differs from [32] where only the relay selection was studied for EH DF relay networks without the optimization of power splitting. Typically, a higher PSR means that more harvested energy is used to transmit the decoded outcome of a source signal, which, however, results in less energy left at a DF relay to decode the source signal along with more decoding errors occurred, implying a tradeoff between the transmission energy and decoding energy.}} Second, we derive closed-form expressions of the outage probability for both the proposed OPS-RS and conventional equal power splitting based relay selection (EPS-RS). Third, the diversity gains of both OPS-RS and EPS-RS are characterized through an asymptotic outage probability analysis in the high signal-to-noise ratio (SNR) region. Finally, we examine an extension of our OPS-RS framework to an energy-harvesting battery (EHB) enabled cooperative relay scenario, and propose an AF based EHB-OPS-RS and a DF based EHB-OPS-RS schemes for AF and DF relay networks, respectively.

The rest of this paper is organized as follows. In Section II, we describe the system model of a cooperative EH relay network. Section III presents the OPS-RS and EPS-RS as well as their outage probability analysis. In Section IV, the diversity analysis of OPS-RS and EPS-RS is carried out. Next, an extension of our OPS-RS framework to an EHB enabled cooperative relay scenario is considered in Section V, where an AF and a DF based EHB-OPS-RS schemes are proposed for AF and DF relay networks, respectively. Section VI provides some numerical outage probability results of the OPS-RS, EPS-RS as well as the AF and DF based EHB-OPS-RS schemes. Finally, Section {{VII}} gives some concluding remarks.

\section{System Model}

\begin{figure}
\centering
\includegraphics[scale=0.55]{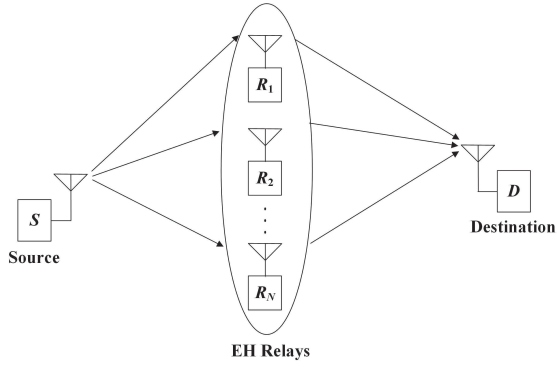}\caption{A cooperative EH relay system consisting of one source (S), one destination (D) and ${N}$ EH relays.}
\end{figure}

As shown in Fig. 1, we consider an EH relay system consisting of a source ($S$), a destination ($D$) and ${N}$ {{separate}} EH relay nodes denoted by ${\mathop R\nolimits_i }$, ${i \in \left\{ 1,2,\ldots,N \right\}}$ {{each equipped with a single antenna}}, where the destination is out of the transmit coverage of source node and an EH relay is opportunistically chosen to assist the source-destination transmission. Notice that all the relays are assumed to harvest energies from their received source signals without stable power supply. {{Moreover, the EH relays can be either dedicated nodes deployed for helping the source transmit to the destination or peer nodes that assist the source-destination transmission and obtain incentives in terms of energy and spectrum resources.}} Fig. 2 shows a block diagram for an EH relay $R_i$, where a received source signal is first divided with a power splitter (PS) into two separate parts, which are respectively fed to the signal decoder for decoding the source message and the energy harvester for supplying the power of subsequent information relaying module. {{It can be observed that compared to the TS strategy, the PS protocol consumes no extra time resources, which is thus used throughout this paper.}} Although only the PS protocol is considered, similar performance analysis and results could be obtained for the TS protocol.

\begin{figure}
\centering
\includegraphics[scale=0.55]{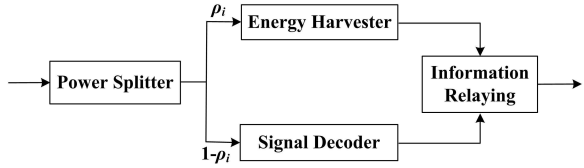}\caption{A block diagram of an EH relay $R_i$.}
\end{figure}

Without loss of generality, let $P$ and $R$ denote the transmit power and data rate of source node, respectively. The PSR of EH relay $R_i$ is represented by ${{\rho _i}}$ (${0 \le {\rho _i} \le 1}$), which is defined as the ratio of {{the}} energy harvested for the information relaying to the total energy received at the relay $R_i$. The remaining fraction $1 - \rho_i$ of the total received energy is used for the information decoder. Moreover, fading coefficients of the channel from the source to EH relay $R_i$ and that from $R_i$ to destination are denoted by ${{ {{h_{si}}}}}$ and ${{ {{h_{id}}} }}$, respectively. In addition, a zero-mean additive white Gaussian noise (AWGN) with a variance of $N_0$ is assumed at any receiver of Fig. 1.

Following the existing literature on EH communications (e.g., [26], [32]-[34]), the Rayleigh fading is used to model the fading coefficients of ${{ {{h_{si}}}}}$ and ${{ {{h_{id}}} }}$. Although the Rayleigh fading is used in this paper, similar {{outage probability analysis}} can be obtained for other fading models (e.g. Rician fading) {{by simply modeling ${{ {{h_{si}}}}}$ and ${{ {{h_{id}}} }}$ with the corresponding probability density function instead (e.g., Rician distribution)}}. Accordingly, one can readily obtain that ${{\left| {{h_{si}}} \right|^2}}$ and ${{\left| {{h_{id}}} \right|^2}}$ follow exponential distributions, whose probability distribution functions can be expressed as
\begin{equation}\label{equa1}
{\mathop f\nolimits_{{{\left| {{h_{si}}} \right|}^2}} \left( x \right) = \frac{1}{{\mathop \sigma \nolimits_{si}^2 }}\exp ( { - \frac{x}{{\mathop \sigma \nolimits_{si}^2 }}} )},
\end{equation}
and
\begin{equation}\label{equa2}
{\mathop f\nolimits_{{{\left| {{h_{id}}} \right|}^2}} \left( x \right) = \frac{1}{{\mathop \sigma \nolimits_{id}^2 }}\exp ( { - \frac{x}{{\mathop \sigma \nolimits_{id}^2 }}} )},
\end{equation}
where ${\mathop \sigma \nolimits_{{{s}}i}^2 }$ and ${\mathop \sigma \nolimits_{{{i}}d}^2 }$ are the means of random variables ${{\left| {{h_{si}}} \right|^2}}$ and ${{\left| {{h_{id}}} \right|^2}}$, respectively.

Considering that the source node broadcasts its signal ${{x_s}}$ (${E( {\mathop {\left| {{x_s}} \right|}\nolimits^2 } ) = 1}$) with a transmit power of ${P}$ to the ${N}$ EH relays, we can express the signal received at an EH relay ${\mathop R\nolimits_i }$ as
\begin{equation}\label{equa3}
{{y_{si}} = {h_{si}}\sqrt P {x_s} + {n_{si}}},
\end{equation}
where ${{n_{si}}}$ is a zero-mean additive white Gaussian noise (AWGN) with a variance of ${\mathop N\nolimits_0 }$ encountered at the EH relay ${\mathop R\nolimits_i }$. As aforementioned, the relay ${\mathop R\nolimits_i }$ first divides its received signal with a power splitter into two separate components for the information decoder and energy harvester, respectively. To be specific, a fraction $ {{\rho _i}}$ of the total received energy is allocated for the energy harvester to supply the power of information relaying module and the remaining fraction $1 - \rho_i$ is used for the information decoder to decode the source message. As a consequence, the energy collected at the energy harvester of relay $R_i$ can be obtained from (3) as
\begin{equation}\label{equa4}
{{E_i} = {\rho _i}\eta P{\left| {{h_{si}}} \right|^2}T},
\end{equation}
where $\eta$ is a conversion efficiency of the energy harvester and $T$ represents the duration of a time slot. Following the existing literature on EH communications [26]-[41], we consider the use of a linear EH model with the perfect channel state information (CSI) available, as given by (4). {{Although a non-linear EH model proposed in [42] is more general in practice, it is analytically untractable as discussed in [43]. To this end, the linear EH model is often assumed along with the perfect CSI available for the purpose of tractability, which has been widely adopted in the existing literature [26]-[41] and [43]. It is indeed interesting to explore an extension to a general scenario with CSI errors and nonlinear EH model, which is out of the scope of this paper and considered for future work.}} Hence, the transmit power used in the information relaying module of relay ${\mathop R\nolimits_i }$ during the following time slot with a duration of ${T}$ can be expressed as
\begin{equation}\label{equa5}
{{P_i}^{t} = {\rho _i}\eta P{\left| {{h_{si}}} \right|^2}}.
\end{equation}

As mentioned above, the relay $R_i$ utilizes the remaining fraction $1 - \rho_i$ of the total received energy for decoding the source message $x_s$. {{Noting that the remaining part of the received signal $y_{si}$ is directly employed for decoding $x_s$ without the energy conversion, we thus express the received signal power for information decoder at the relay $R_i$ as}}
\begin{equation}\label{equa6}
{{P_i}^{d} = (1 - {\rho _i})P{\left| {{h_{si}}} \right|^2}},
\end{equation}
from which the received signal-to-noise ratio (SNR) at the relay ${\mathop R\nolimits_i }$ is obtained as
\begin{equation}\label{equa7}
{{\gamma _{si}} = (1 - {\rho _i})\gamma {\left| {{h_{si}}} \right|^2}},
\end{equation}
where ${\gamma  = {P}/{{{N_0}}}}$. From (7), the channel capacity from the source to relay ${\mathop R\nolimits_i }$ can be given by
\begin{equation}\label{equa8}
{{C_{si}} = \frac{1}{2}\log [ {1 + (1 - {\rho _i})\gamma {{\left| {{h_{si}}} \right|}^2}} ]}.
\end{equation}

Next, among the $N$ EH relays, an EH relay that succeeds in decoding the source message $x_s$ is opportunistically chosen to forward its decode result in another time slot. Without loss of generality, let us consider that the EH relay $R_i$ succeeding in decoding $x_s$ is selected to forward the source message to the destination with a transmit power of ${P_i}^{t}$ as given by (5). Thus, the corresponding received signal at the destination is written as
\begin{equation}\label{equa9}
\begin{split}
{y_{id}} = {h_{id}}\sqrt {{P_i}^{t}} {x_s} + {n_{id}}= {h_{id}}\sqrt {{\rho _i}\eta P{\left| {{h_{si}}} \right|^2}} {x_s} + {n_{id}}.
\end{split}
\end{equation}
where ${{n_{id}}}$ is a zero-mean AWGN with a variance of ${\mathop N\nolimits_0 }$ encountered at the destination. From (9), we can readily obtain the received SNR at the destination as
\begin{equation}\label{equa10}
{{\gamma _{id}} = {\rho _i}\gamma \eta {\left| {{h_{si}}} \right|^2}{\left| {{h_{id}}} \right|^2}},
\end{equation}
from which the channel capacity from the relay ${\mathop R\nolimits_i }$ and destination is given by
\begin{equation}\label{equa11}
{{C_{id}} = \frac{1}{2}\log ( {1 + {\rho _i}\gamma \eta {{\left| {{h_{si}}} \right|}^2}{{\left| {{h_{id}}} \right|}^2}} )}.
\end{equation}

Following existing literature on DF relaying transmissions [1]-[3], an overall channel capacity from the source to destination via a relay is given by the minimum of the channel capacity from source to relay and that from relay to destination. Hence, using (8) and (11), the overall channel capacity from the source via an EH relay $R_i$ to destination can be expressed as
\begin{equation}\label{equa12}
{\mathop C \nolimits_{sid} = \mathop {\min }\limits_{} ({C _{si}},{C _{id}})},
\end{equation}
which completes the system model of EH relay transmissions with the DF protocol.

\section{Joint Power Splitting and Relay Selection}
In this section, we first propose the OPS-RS scheme for cooperative EH relay networks. For comparison purposes, the traditional EPS-RS is also considered as a baseline. Closed-form outage probability expressions are derived for the EPS-RS and OPS-RS schemes over Rayleigh fading channels.

\subsection{EPS-RS Scheme}
This subsection presents the conventional EPS-RS scheme as a benchmark. As aforementioned, the source first transmits its signal $x_s$ to the EH relays which then divide their received radio signals with a power-splitter into two separate components for the energy harvester and information decoder, respectively. As implied in the name, the EPS-RS scheme allows an EH relay to divide the total energy of its received signal into two equal parts, i.e., the PSR of each EH relay is given by ${\mathop \rho \nolimits_{i}^{{\textrm{EPS}}} = {1}/{2}}$ for the EPS-RS scheme. In other words, one half of the total received power is used for the energy harvester and the remaining power is used for the information decoder. Substituting ${\mathop \rho \nolimits_{i}^{{\textrm{EPS}}} = {1}/{2}}$ into (8), we can obtain the channel capacity from the source to EH relay ${\mathop R\nolimits_i }$ for the EPS-RS scheme as
\begin{equation}\label{equa13}
{\mathop C\nolimits_{si}^{\textrm{EPS}}  = \frac{1}{2}\log ( {1 + \frac{1}{2}\gamma {{\left| {{h_{si}}} \right|}^2}} )}.
\end{equation}
Similarly, given ${\mathop \rho \nolimits_{i}^{{\textrm{EPS}}} = {1}/{2}}$, the channel capacity from the EH relay ${\mathop R\nolimits_i }$ to destination for the EPS-RS scheme can be obtained from (11) as
\begin{equation}\label{equa14}
{\mathop C\nolimits_{id}^{\textrm{EPS}}  = \frac{1}{2}\log ( {1 + \frac{1}{2}\gamma \eta {{\left| {{h_{si}}} \right|}^2}{{\left| {{h_{id}}} \right|}^2}} )}.
\end{equation}
From (12)-(14), we can obtain an overall channel capacity from the source via an EH relay $R_i$ to destination for the EPS-RS scheme as
\begin{equation}\label{equa15}
{C^{\textrm{EPS}}_{sid} = \mathop {\min }\limits_{} ({C^{\textrm{EPS}} _{si}},{C^{\textrm{EPS}} _{id}})},
\end{equation}
where ${C^{\textrm{EPS}} _{si}}$ and ${C^{\textrm{EPS}} _{id}}$ are given by (13) and (14), respectively.

Typically, an EH relay that maximizes the overall channel capacity from the source to destination is opportunistically selected to assist the source-destination transmission. Therefore, from (15), an opportunistic relay selection criterion can be written as
\begin{equation}\label{equa16}
\begin{split}
\mathop R\nolimits_b^{\textrm{EPS}}  &= \arg \;\mathop {\max }\limits_{i = 1, 2, \cdots ,N} \;\mathop {\min }\limits_{} (\mathop C\nolimits_{si}^{\textrm{EPS}} ,\mathop C\nolimits_{id}^{\textrm{EPS}} )\\
&=\arg \mathop {\max }\limits_{i = 1, 2, \cdots ,N} \min (|h_{si}|^2,\eta |h_{si}|^2 |{h_{id}}{|^2}),
\end{split}
\end{equation}
where $\mathop R\nolimits_b^{\textrm{EPS}} $ denotes the best EH relay selected by the EPS-RS scheme. According to (16), the overall channel capacity from the source to destination for the EPS-RS scheme is given by
\begin{equation}\label{equa17}
{ C_{sd}^{\textrm{EPS}} = \mathop {\max }\limits_{i = 1, 2, \cdots ,N} \mathop {\min }\limits_{} ({C^{\textrm{EPS}} _{si}},{C^{\textrm{EPS}} _{id}})}.
\end{equation}

In what follows, we present an outage probability analysis for the conventional EPS-RS scheme. As is known, an outage event happens when the channel capacity falls below a predefined data rate. Thus, the probability of occurrence of an outage event (also called outage probability) for the conventional EPS-RS scheme is obtained as
\begin{equation}\label{equa18}
P_{{\rm{out}}}^{{\textrm{EPS}}} = \Pr ({C_{sd}^{{\textrm{EPS}}}} < R),
\end{equation}
where $R$ denotes the data rate of source-destination transmissions. Substituting ${C_{sd}^{{\textrm{EPS}}}}$ from (17) into (18) {{and noting that $C^{\textrm{EPS}}_{sid}$ for different relays ($i=1,2,\cdots,N$) are independent of each other}}, we have
\begin{equation}\label{equa19}
\begin{split}
\mathop P\nolimits_{{\rm{out}}}^{{\textrm{EPS}}}  &= \Pr \left[ {\mathop {\max }\limits_{i = 1, 2, \cdots ,N} \;\mathop {\min }\limits_{} \left(\mathop C\nolimits_{si}^{\textrm{EPS}} ,\mathop C\nolimits_{id}^{\textrm{EPS}} \right) < R} \right]\\
&= \prod\limits_{i = 1}^N {\Pr \left[ {\min \left( {{{\left| {{h_{si}}} \right|}^2},\eta {{\left| {{h_{si}}} \right|}^2}{{\left| {{h_{id}}} \right|}^2}} \right) < \beta } \right]} \\
&= \prod\limits_{i = 1}^N {\left[ {1 - \underbrace {\Pr \left( {{{\left| {{h_{si}}} \right|}^2} > \beta ,{{\left| {{h_{id}}} \right|}^2} > \frac{\alpha }{{{{\left| {{h_{si}}} \right|}^2}}}} \right)}_{\Phi_i}} \right]},
\end{split}
\end{equation}
where ${\alpha  = {{2( {\mathop 2\nolimits^{2R}  - 1})} \mathord{\left/
 {\vphantom {{2 \cdot \left( {\mathop 2\nolimits^{2R}  - 1} \right)} {(\gamma \eta )}}} \right.
 \kern-\nulldelimiterspace} {(\gamma \eta )}}}$ and ${\beta  = {{2 ( {\mathop 2\nolimits^{2R}  - 1})} \mathord{\left/
 {\vphantom {{2 \cdot \left( {\mathop 2\nolimits^{2R}  - 1} \right)} \gamma }} \right.
 \kern-\nulldelimiterspace} \gamma }}$. Noting that ${{\left| {{h_{si}}} \right|^2}}$ and ${{\left| {{h_{id}}} \right|^2}}$ are independent exponentially distributed with respective means of ${\mathop \sigma \nolimits_{{{s}}i}^2 }$ and ${\mathop \sigma \nolimits_{{{i}}d}^2 }$, and denoting ${\left| {{h_{si}}} \right|^2} = X$ and ${{\left| {{h_{id}}} \right|^2} = Y}$, we can rewrite the term ${\Phi_i}$ of (19) as
\begin{equation}\label{equa20}
\begin{split}
{\Phi _i} &= \Pr \left( {X > \beta ,Y > \frac{\alpha }{X}} \right) \\
&= \frac{1}{{\sigma _{si}^2}}\int_\beta ^\infty  {\exp ( - \frac{x}{{\sigma _{si}^2}})\exp ( - \frac{\alpha }{{\sigma _{id}^2x}})dx}.  \\
\end{split}
\end{equation}
Using the Maclaurin series expansion, we have
\begin{equation}\label{equa21}
\exp ( - \frac{\alpha }{{\sigma _{id}^2x}}) = \sum\limits_{u = 0}^\infty  {\frac{{{{( - 1)}^u}{a^u}}}{{u!\sigma _{id}^{2u}{x^u}}}},
\end{equation}
for $-\infty < x < \infty$. Combining (20) and (21), we arrive at
$${\Phi _i} = \frac{1}{{\sigma _{si}^2}}\sum\limits_{u = 0}^\infty  {\int_\beta ^\infty  {\frac{{{{( - 1)}^u}{a^u}}}{{u!\sigma _{id}^{2u}{x^u}}}\exp ( - \frac{x}{{\sigma _{si}^2}})dx} },$$
which can be expanded to
\begin{equation}\label{equa22}
\begin{split}
 {\Phi _i} &= \frac{1}{{\sigma _{si}^2}}\int_\beta ^\infty  {\exp ( - \frac{x}{{\sigma _{si}^2}})dx}  - \frac{a}{{\sigma _{si}^2\sigma _{id}^2}}\int_\beta ^\infty  {\frac{1}{x}\exp ( - \frac{x}{{\sigma _{si}^2}})dx}  \\
 &\quad+ \frac{1}{{\sigma _{si}^2}}\sum\limits_{u = 2}^\infty  {\int_\beta ^\infty  {\frac{{{{( - 1)}^u}{a^u}}}{{u!\sigma _{id}^{2u}{x^u}}}\exp ( - \frac{x}{{\sigma _{si}^2}})dx} }  \\
  &= \exp ( - \frac{\beta }{{\sigma _{si}^2}}) - \frac{a}{{\sigma _{si}^2\sigma _{id}^2}}Ei(\frac{\beta }{{\sigma _{si}^2}}) + \frac{1}{{\sigma _{si}^2}}\sum\limits_{u = 2}^\infty  {\frac{{{{( - 1)}^u}{a^u}}}{{u!\sigma _{id}^{2u}}}{\Phi _{i,u}}},  \\
\end{split}
\end{equation}
where ${Ei\left( x  \right) = \int_x ^\infty  {\frac{{\mathop e\nolimits^{ - t} }}{t}} dt}$ is known as the exponential integral function and ${\Phi _{i,u}}$ is given by
\begin{equation}\nonumber
\begin{split}
{\Phi _{i,u}} = &\int_\beta ^\infty  {\frac{1}{{{x^u}}}\exp ( - \frac{x}{{\sigma _{si}^2}})dx}  \\
=& \exp ( - \frac{\beta }{{\sigma _{si}^2}})\sum\limits_{v = 1}^{u - 1} {\frac{{(v - 1)!{{( - 1)}^{u - v - 1}}}}{{(u - 1)!{\beta ^v}\sigma _{si}^{2(u - v - 1)}}}} \\
 &- \frac{{{{( - 1)}^{u - 1}}}}{{(u - 1)!\sigma _{si}^{2(u - 1)}}}Ei( - \frac{\beta }{{\sigma _{si}^2}}). \\
\end{split}
\end{equation}

Finally, substituting ${\Phi_i}$ from (22) into (19) yields
\begin{equation}\label{equa23}
P_{\rm{out}}^{{\textrm{EPS}}} = \prod\limits_{i = 1}^N {\left[ \begin{split}
 &1 - \exp ( - \frac{\beta }{{\sigma _{si}^2}}) + \frac{a}{{\sigma _{si}^2\sigma _{id}^2}}Ei(\frac{\beta }{{\sigma _{si}^2}}) \\
 &- \frac{1}{{\sigma _{si}^2}}\sum\limits_{u = 2}^\infty  {\frac{{{{( - 1)}^u}{a^u}}}{{u!\sigma _{id}^{2u}}}{\Phi _{i,u}}}  \\
 \end{split} \right],}
\end{equation}
which gives a closed-form expression of outage probability for the EPS-RS scheme. {{It needs to be pointed out that an infinite series is involved in (23) due to the use of Maclaurin series expansion as given by (21), which can be converged quickly after a few iterations of $u$.}}

\subsection{OPS-RS Scheme}
In this section, we propose the OPS-RS scheme and derive its closed-form outage probability over Rayleigh fading channels. In the proposed OPS-RS scheme, the source first broadcasts its message $x_s$ to $N$ EH relays which then perform an optimal power splitting for the sake of maximizing the overall channel capacity from source to destination. Finally, an EH relay with an optimized power splitter that has the highest overall channel capacity is chosen to assist the source transmission to destination. This differs from the conventional EPS-RS approach, where the total energy received at an EH relay is simply divided into two equal parts. Without loss of generality, let us consider that the source transmits its signal to the destination via an EH relay $R_i$ with a PSR $\rho_i$. Hence, from (12), an optimal PSR $\rho^{\textrm{OPS}}_i$ for maximizing the overall channel capacity from source via $R_i$ to destination can be obtained as
\begin{equation}\label{equa24}
\rho^{\textrm{OPS}}_i  = \arg \mathop {\max }\limits_{0 \le {\rho _i} \le 1} \;{C_{sid}} = \mathop {\max }\limits_{0 \le {\rho _i} \le 1} \;\mathop {\min }\limits_{} ({C_{si}},{C_{id}}),
\end{equation}
where $C_{si}$ and $C_{id}$ are given by (8) and (11), respectively. Combining (8), (11) and (24), we can further obtain
\begin{equation}\label{equa25}
\rho^{\textrm{OPS}}_i  = \arg  \mathop {\max }\limits_{0 \le {\rho _i} \le 1} \;\mathop {\min }\limits_{} (1 - \mathop \rho \nolimits_i ,{\rho _i}\eta {\left| {{h_{id}}} \right|^2}).
\end{equation}
{{It can be observed from (25) that if the term $1 - \mathop \rho \nolimits_i $ is not equal to ${\rho _i}\eta {\left| {{h_{id}}} \right|^2}$, a higher value of $\mathop {\min }\limits_{} (1 - \mathop \rho \nolimits_i ,{\rho _i}\eta {\left| {{h_{id}}} \right|^2})$ can always be obtained by either increasing $\rho _i$ to increase ${\rho _i}\eta {\left| {{h_{id}}} \right|^2}$ and decrease $1 - \mathop \rho \nolimits_i$ or decreasing $\rho _i$ to increase $1 - \mathop \rho \nolimits_i$ and decrease ${\rho _i}\eta {\left| {{h_{id}}} \right|^2}$, until $1 - \mathop \rho \nolimits_i $ becomes the same as ${\rho _i}\eta {\left| {{h_{id}}} \right|^2}$, since they are an increasing and a decreasing functions of $\rho _i$, respectively. Therefore, an optimal PSR $\rho^{\textrm{OPS}}_i$ can be obtained from (25) as}}
\begin{equation}\label{equa26}
1 - \rho^{\textrm{OPS}}_i  = \rho^{\textrm{OPS}}_i\eta {\left| {{h_{id}}} \right|^2},
\end{equation}
which, in turn, leads to
\begin{equation}\label{equa27}
\rho^{\textrm{OPS}}_i = \frac{1}{{1 + \eta {{\left| {{h_{id}}} \right|}^2}}}.
\end{equation}
Substituting the optimal PSR $\rho^{\textrm{OPS}}_i$ from (27) into (12), we can obtain an overall channel capacity from the source to destination via EH relay $R_i$ with an optimized power-splitter for the OPS-RS scheme as
\begin{equation}\label{equa28}
\mathop C\nolimits_{sid}^{\textrm{OPS}}  = \frac{1}{2}\log (1 + \frac{{\gamma \eta {{\left| {{h_{si}}} \right|}^2}{{\left| {{h_{id}}} \right|}^2}}}{{1 + \eta {{\left| {{h_{id}}} \right|}^2}}}).
\end{equation}

In the OPS-RS scheme, an EH relay having the highest overall channel capacity of $\mathop C\nolimits_{sid}^{\textrm{OPS}}$ is chosen to forward the source message to destination. Hence, using (28), we can obtain a relay selection criterion for the OPS-RS scheme as
\begin{equation}\label{equa29}
\begin{split}
\mathop R\nolimits_b^{\textrm{OPS}}  &= \arg \;\mathop {\max }\limits_{i = 1, 2, \cdots ,N} \;\mathop C\nolimits_{sid}^{\textrm{OPS}}\\
&=\arg \mathop {\max }\limits_{i = 1, 2, \cdots ,N} \frac{{|{h_{si}}{|^2}|{h_{id}}{|^2}}}{{1 + \eta |{h_{id}}{|^2}}},
\end{split}
\end{equation}
where $\mathop R\nolimits_b^{\textrm{OPS}} $ denotes the best EH relay selected by the OPS-RS scheme. As shown in (27) and (29), the channel fading amplitudes of $|h_{si}|^2$ and $|h_{id}|^2$ are needed at an EH relay $R_i$ to perform the optimization of PSR $\rho_i$ and relay selection. It is pointed out that the fading amplitude of $|h_{si}|^2$ may be obtained at the relay $R_i$ through channel estimation. Moreover, the destination can estimate the fading amplitude of $|h_{id}|^2$ and then feed it back to the relay $R_i$. In addition, a distributed relay selection framework can be implemented at the separate EH relays. {{More specifically, each EH relay maintains a timer whose initial value is set in inverse proportional to the term ${{|{h_{si}}{|^2}|{h_{id}}{|^2}}}/({{1 + \eta |{h_{id}}{|^2}}})$ as given by (29). In this way, the EH relay with the smallest initial value becomes the best relay node, which exhausts its timer earliest and then broadcasts a control packet to notify the source, destination and other EH relays [15].}} From (29), the overall source-destination channel capacity for the OPS-RS scheme is given by
\begin{equation}\label{equa30}
{C_{sd}^{\textrm{OPS}}} = \mathop {\max }\limits_{i = 1, 2, \cdots ,N} C_{sid}^{\textrm{OPS}},
\end{equation}
where $C_{sid}^{\textrm{OPS}}$ is given by (28).

The following presents an outage probability analysis for the proposed OPS-RS scheme. As discussed, an outage event occurs when the channel capacity of proposed OPS-RS scheme ${C_{sd}^{\textrm{OPS}}}$ drops below a predefined data rate $R$. Hence, using (30) {{and noting that $C^{\textrm{OPS}}_{sid}$ for different relays ($i=1,2,\cdots,N$) are independent of each other}}, we can obtain an outage probability of the OPS-RS scheme as
\begin{equation}\label{equa31}
\begin{split}
\mathop P\nolimits_{{\rm{out}}}^{{\textrm{OPS}}}  &  = \Pr \left( {{C_{sd}^{{\textrm{OPS}}}} < R} \right)\\
&= \Pr \left( {\mathop {\max }\limits_{i = 1, 2, \cdots ,N} \;\mathop C\nolimits_{sid}^{\textrm{OPS}}  < R} \right) \\
&= \prod\limits_{i = 1}^N {{\Pr \left( {C_{sid}^{{\textrm{OPS}}} < R} \right)}}.
\end{split}
\end{equation}
Substituting $C_{sid}^{{\textrm{OPS}}}$ from (28) into (31) gives
\begin{equation}\label{equa32}
\mathop P\nolimits_{{\rm{out}}}^{{\textrm{OPS}}} = \prod\limits_{i = 1}^N {\Pr \left( {\frac{1}{2}\log (1 + \frac{{\gamma \eta {{\left| {{h_{si}}} \right|}^2}{{\left| {{h_{id}}} \right|}^2}}}{{1 + \eta {{\left| {{h_{id}}} \right|}^2}}}) <R} \right)} ,
\end{equation}
which is further simplified to
\begin{equation}\label{equa33}
\begin{split}
\mathop P\nolimits_{{\rm{out}}}^{{\textrm{OPS}}} =& \prod\limits_{i = 1}^N {\Pr \left( {|{h_{si}}{|^2}|{h_{id}}{|^2} < \frac{{{2^{2R}} - 1}}{{\gamma \eta }}(1 + \eta |{h_{id}}{|^2})} \right)} \\
=&\prod\limits_{i = 1}^N {{\Pr \left( {{{\left| {{h_{si}}} \right|}^2} < \frac{{\delta }}{{{{\left| {{h_{id}}} \right|}^2}}} + \delta \eta } \right)}},
\end{split}
\end{equation}
where ${\delta  = \frac{\mathop 2\nolimits^{2R}  - 1 }{{\gamma \eta }}}$. Again, noting that random variables ${{\left| {{h_{si}}} \right|^2}}$ and ${{\left| {{h_{id}}} \right|^2}}$ are independent exponentially distributed with respective means of ${\mathop \sigma \nolimits_{{{s}}i}^2 }$ and ${\mathop \sigma \nolimits_{{{i}}d}^2 }$, and denoting ${\left| {{h_{si}}} \right|^2} = X$ and ${{\left| {{h_{id}}} \right|^2} = Y}$, we can rewrite (33) as
\begin{equation}\label{equa34}
P_{\rm{out}}^{{\textrm{OPS}}} = \prod\limits_{i = 1}^N {\left[ {\int_0^\infty  {\frac{1}{{\sigma _{id}^2}}\exp ( - \frac{y}{{\sigma _{id}^2}})[1 - \exp ( - \frac{\delta }{{\sigma _{si}^2y}} - \frac{{\delta \eta }}{{\sigma _{si}^2}})]dy} } \right]},
\end{equation}
which is further given by
\begin{equation}\label{equa35}
\begin{split}
 P_{\rm{out}}^{{\textrm{OPS}}} & = \prod\limits_{i = 1}^N {\left[ {1 - \exp ( - \frac{{\delta \eta }}{{\sigma _{si}^2}})\int_0^\infty  {\frac{1}{{\sigma _{id}^2}}\exp ( - \frac{y}{{\sigma _{id}^2}} - \frac{\delta }{{\sigma _{si}^2y}})dy} } \right]}  \\
&= \prod\limits_{i = 1}^N {\left[ {1 - \frac{{2\sqrt \delta  }}{{{\sigma _{si}}{\sigma _{id}}}}\exp ( - \frac{{\delta \eta }}{{\sigma _{si}^2}}){K_1}(\frac{{2\sqrt \delta  }}{{{\sigma _{si}}{\sigma _{id}}}})} \right]},  \\
 \end{split}
\end{equation}
where ${\mathop K\nolimits_1 \left(  \cdot  \right)}$ is the first-order modified Bessel function of second kind as given by (8.432.6) {{in}} [44]. So far, we have derived closed-form outage probability expressions for both the EPS-RS and OPS-RS schemes over Rayleigh fading channels.

\section{Diversity Gain Analysis}
In this section, we present an asymptotic outage probability analysis in the high SNR region to characterize diversity gains of the EPS-RS and OPS-RS schemes. Following [45] and [46], the diversity gain is defined as a ratio of the logarithm of an outage probability to the logarithm of SNR as the SNR approaches to infinity, namely
\begin{equation}\label{equa36}
{d =  - \mathop {\lim }\limits_{\gamma  \to \infty } \frac{{\log \mathop P\nolimits_{\rm{out}} \left( \gamma  \right)}}{{\log \gamma }}},
\end{equation}
where ${\mathop P\nolimits_{\rm{out}} \left( \gamma  \right)}$ represents an outage probability as a function of the SNR ${\gamma }$.

\subsection{EPS-RS}
This subsection conducts the diversity analysis of traditional EPS-RS scheme. From (36), the diversity gain of EPS-RS scheme can be obtained as
\begin{equation}\label{equa37}
{\mathop d\nolimits_{{\textrm{EPS}}} =  - \mathop {\lim }\limits_{\gamma  \to \infty } \frac{{\log \left( {\mathop P\nolimits_{{\rm{out}}}^{{\textrm{EPS}}} } \right)}}{{\log \gamma }}},
\end{equation}
where ${\mathop P\nolimits_{{\rm{out}}}^{{\textrm{EPS}}} }$ is given by (19). Substituting $\mathop \Phi \nolimits_1$ from (20) into (19) gives
\begin{equation}\label{equa38}
P_{\rm{out}}^{{\textrm{EPS}}} = \prod\limits_{i = 1}^N {\left( {1 - \frac{1}{{\sigma _{si}^2}}\int_\beta ^\infty  {\exp ( - \frac{x}{{\sigma _{si}^2}} - \frac{\alpha }{{\sigma _{id}^2x}})dx} } \right)},
\end{equation}
where $\alpha  = \frac{{2({2^{2R}} - 1)}}{{\eta \gamma }}$ and $\beta  = \frac{{2({2^{2R}} - 1)}}{\gamma }$.

\textbf{Proposition 1:} \emph{Given an exponential random variable ${x}$ ($x>\beta$) with a mean of ${\mathop \sigma \nolimits_{si}^2 }$, the following equation holds with one probability
$${\frac{{ \alpha }}{\sigma^2_{id}x} = 0},$$
for ${\gamma  \to \infty }$, where ${\alpha  = {{2( {\mathop 2\nolimits^{2R}  - 1})} \mathord{\left/
 {\vphantom {{2 \cdot \left( {\mathop 2\nolimits^{2R}  - 1} \right)} {(\gamma \eta )}}} \right.
 \kern-\nulldelimiterspace} {(\gamma \eta )}}}$ and $\beta  = {{2({2^{2R}} - 1)}}/{\gamma }$.}

\textbf{Proof:} See Appendix A for details. {{It is pointed out that the term ${{ \alpha }}/({\sigma^2_{id}x})$ becomes random due to the presence of an exponential random variable $x$ in Proposition 1. Moreover, the expectation of ${1}/{x}$ ($x>\beta$) is given by $\frac{1}{\sigma^2_{si}}Ei(\frac{\beta}{\sigma^2_{si}})$ that tends to infinity for $\gamma \to \infty$, resulting in an uncertainty of the convergence of the random term ${{ \alpha }}/({\sigma^2_{id}x})$, which motivates the proof of Proposition 1.}}

Using Proposition 1, we can obtain $P_{\rm{out}}^{{\textrm{EPS}}}$ from (38) as
\begin{equation}\label{equa39}
\begin{split}
P_{\rm{out}}^{{\textrm{EPS}}} & = \prod\limits_{i = 1}^N {\left[ {1 - \frac{1}{{\sigma _{si}^2}}\int_\beta ^\infty  {\exp ( - \frac{x}{{\sigma _{si}^2}})dx} } \right]}\\
&= \prod\limits_{i = 1}^N {\left( {1 - \exp ( - \frac{\beta }{{\sigma _{si}^2}})} \right)},
\end{split}
\end{equation}
for ${\gamma  \to \infty }$. Noting $\beta  = \frac{{2({2^{2R}} - 1)}}{\gamma }$ and using the Taylor expansion for ${\gamma  \to \infty }$, we can arrive at
\begin{equation}\label{equa40}
1 - \exp ( - \frac{\beta }{{\sigma _{si}^2}}) = \frac{\beta }{{\sigma _{si}^2}},
\end{equation}
where higher-order infinitesimals are ignored. Combining (39) and (40), we obtain
\begin{equation}\label{equa41}
P_{\rm{out}}^{{\textrm{EPS}}} = \prod\limits_{i = 1}^N {\frac{{2({2^{2R}} - 1)}}{{\sigma _{si}^2}}}  \cdot \frac{1}{{{\gamma ^N}}},
\end{equation}
for ${\gamma  \to \infty }$. Substituting $P_{\rm{out}}^{{\textrm{EPS}}}$ from (41) into (37), the diversity gain of EPS-RS scheme is given by
\begin{equation}\label{equa42}
d_{{\textrm{EPS}}}  = N,
\end{equation}
which shows that a diversity gain of $N$ is achieved by the traditional EPS-RS scheme.

\subsection{OPS-RS}
In this subsection, we analyze the diversity gain of proposed OPS-RS scheme. From (30), the diversity of OPS-RS scheme can be obtained as
\begin{equation}\label{equa43}
{\mathop d\nolimits_{{\textrm{OPS}}} =  - \mathop {\lim }\limits_{\gamma  \to \infty } \frac{{\log \left( {\mathop P\nolimits_{{\rm{out}}}^{{\textrm{OPS}}} } \right)}}{{\log \gamma }}},
\end{equation}
where $P_{{\rm{out}}}^{{\textrm{OPS}}}$ is given by (35).
Noting $\delta  = \frac{{{2^{2R}} - 1}}{{\eta \gamma }}$ and letting ${\gamma  \to \infty }$, we can obtain the term $\frac{{2\sqrt \delta  }}{{\sigma _{si}\sigma _{id}}}$ approaching to zero, namely
$$\frac{{2\sqrt \delta  }}{{{\sigma _{si}}{\sigma _{id}}}} \to 0,$$
which, in turn, leads to
\begin{equation}\label{equa44}
\frac{{2\sqrt \delta  }}{{{\sigma _{si}}{\sigma _{id}}}}{K_1}(\frac{{2\sqrt \delta  }}{{{\sigma _{si}}{\sigma _{id}}}}) = 1.
\end{equation}
Combining (35) and (44), we have
\begin{equation}\label{equa45}
P_{\rm{out}}^{{\textrm{OPS}}} = \prod\limits_{i = 1}^N {\left( {1 - \exp ( - \frac{{\delta \eta }}{{\sigma _{si}^2}})} \right)},
\end{equation}
for ${\gamma  \to \infty }$. Substituting $\delta  = \frac{{{2^{2R}} - 1}}{{\eta \gamma }}$ into the preceding equation gives
\begin{equation}\label{equa46}
P_{\rm{out}}^{{\textrm{OPS}}} = \prod\limits_{i = 1}^N {\left( {1 - \exp ( - \frac{{{2^{2R}} - 1}}{{\sigma _{si}^2\gamma }})} \right)} ,
\end{equation}
for ${\gamma  \to \infty }$. Moreover, using the Taylor expansion and ignoring higher-order infinitesimals, we can obtain
\begin{equation}\label{equa47}
1 - \exp ( - \frac{{{2^{2R}} - 1}}{{\sigma _{si}^2\gamma }}) = \frac{{{2^{2R}} - 1}}{{\sigma _{si}^2\gamma }},
\end{equation}
for ${\gamma  \to \infty }$. Combining (46) and (47), we arrive at
\begin{equation}\label{equa48}
P_{\rm{out}}^{{\textrm{OPS}}} = \prod\limits_{i = 1}^N {\frac{{{2^{2R}} - 1}}{{\sigma _{si}^2}}}  \cdot \frac{1}{{{\gamma ^N}}}.
\end{equation}
for ${\gamma  \to \infty }$. Substituting $P_{\rm{out}}^{{\textrm{OPS}}} $ from (48) into (43) yields
\begin{equation}\label{equa49}
{\mathop d\nolimits_{{\textrm{OPS}}}  = N},
\end{equation}
from which one can observe that the OPS-RS scheme achieves the diversity gain of $N$, where $N$ is the number of EH relays. As a consequence, as the number of EH relays ${N}$ increases, the diversity gain of OPS-RS scheme increases accordingly, demonstrating a significant amount of outage performance improvement achieved through increasing the number of EH relays.

\section{Extension to EHB Enabled Cooperative Relay Systems}
In this section, we are focused on an extension of the aforementioned OPS-RS framework to an EHB enabled cooperative relay scenario, where the EH relays are considered to be equipped with batteries used to store their harvested energies.

\subsection{DF based EHB-OPS-RS Scheme}
{{This subsection proposes an EHB-OPS-RS scheme for DF relay networks, referred to as the DF based EHB-OPS-RS scheme. Let $E_i^s$ denote the energy previously harvested and accumulated in the battery of relay $R_i$, which starts with zero at the very beginning. Moreover, once an EH relay is selected to assist the source-destination transmission, the stored energy in its battery is used to re-transmit the source signal. Denoting the energy currently collected at the energy harvester of relay $R_i$ for the information relaying by $E_i^c$, we can express the total energy available for the relay $R_i$ as
\begin{equation}\label{equa50}
 E_i  =  E_i^s +  E_i^c ,
\end{equation}
where $E_i^c$ is given by
\begin{equation}\label{equa51}
{E_i^c} = {\rho _i}\eta P{\left| {{h_{si}}} \right|^2}T.
\end{equation}
From (50) and (51), the transmit power available at the EH relay $R_i$ can be expressed as
\begin{equation}\label{equa52}
\mathop P\nolimits_i^t  = \dfrac{E_i}{T} = \mathop \rho \nolimits_i \eta P\mathop {\left| {\mathop h\nolimits_{si} } \right|}\nolimits^2  + \mathop P\nolimits_i^s,
\end{equation}
where $\mathop P\nolimits_i^s$ represents the power generated from the energy stored at the battery of $R_i$. Meanwhile, similarly to (6), the remaining fraction $1 - \rho_i$ of the currently harvested energy at the relay $R_i$ is used for the information decoder and the corresponding power is written as
\begin{equation}\label{equa53}
P_i^{d} = (1 - {\rho _i})P{\left| {{h_{si}}} \right|^2},
\end{equation}
from which the channel capacity from the source to relay $R_i$ is given by
\begin{equation}\label{equa54}
{C_{si}} = \frac{1}{2}\log \left[ {1 + (1 - {\rho _i})\gamma {{\left| {{h_{si}}} \right|}^2}} \right],
\end{equation}
where $\gamma=P/N_0$. Moreover, by using (52), the channel capacity from the relay $R_i$ to destination is given by
\begin{equation}\label{equa55}
{C_{id}} = \frac{1}{2}\log \left( {1 + {\rho _i}\gamma \eta {{\left| {{h_{si}}} \right|}^2}{{\left| {{h_{id}}} \right|}^2}{\rm{ + }}\mathop \gamma \nolimits_i^s {{\left| {{h_{id}}} \right|}^2}} \right),
\end{equation}
where $\gamma_i^s = P_i^s/N_0$. Similarly to (24), an optimal PSR for the DF based EHB-OPS-RS scheme can be obtained from (54) and (55) as
\begin{equation}\label{equa56}
\rho _{i,\textrm{DF}}^{\textrm{EHB}} = \arg \mathop {\max }\limits_{0 \le {\rho _i} \le 1} \min \left(\begin{split}& (1 - {\rho _i})\gamma {{\left| {{h_{si}}} \right|}^2},\\
&{\rho _i}\gamma \eta {{\left| {{h_{si}}} \right|}^2}{{\left| {{h_{id}}} \right|}^2}{\rm{ + }}\mathop \gamma \nolimits_i^s {{\left| {{h_{id}}} \right|}^2}
\end{split}\right),
\end{equation}
which leads to
\begin{equation}\label{equa57}
\rho _{i,\textrm{DF}}^{\textrm{EHB}}={\left[ {\frac{{1 - \mathop \gamma \nolimits_i^s |{h_{id}}{|^2}/(\gamma |{h_{si}}{|^2})}}{{1 + \eta {{\left| {{h_{id}}} \right|}^2}}}} \right]^ + },
\end{equation}
where ${[x]^ + } = \max (x,0)$. It can be observed that if no batteries are equipped at the EH relays (i.e., ${\mathop \gamma \nolimits_i^s } = 0$), the optimal PSR $\rho _{i,\textrm{DF}}^{\textrm{EHB}}$ of (57) for the DF based EHB-OPS-RS scheme is degraded to that of (27) for the OPS-RS without the battery. Substituting the optimal PSR $\rho _{i,\textrm{DF}}^{\textrm{EHB}}$ from (57) into (54) and (55), we can obtain an overall channel capacity from the source via an EH relay to the destination as
\begin{equation}\label{equa58}
C_{sd,{\textrm{DF}}}^{{\textrm{EHB}}} = \mathop {\max }\limits_{i = 1, 2, \cdots ,N} \frac{1}{2}\log \left[ {1 + (1 - {\rho _{i,\textrm{DF}}^{\textrm{EHB}}})\gamma {{\left| {{h_{si}}} \right|}^2}} \right].
\end{equation}
{{One can see from (57) that an optimal PSR of $\rho _{i,\textrm{DF}}^{\textrm{EHB}} = 0$ arises given ${1 - P _i^s|{h_{id}}{|^2}/(P |{h_{si}}{|^2})} < 0$, which means that the stored battery power $\mathop P\nolimits_i^s$ is sufficiently high for retransmitting the source signal without the need of any additionally harvested energy. In this case, all the currently harvested power $P|h_{si}|^2$ as implied from (53) by substituting $\rho _{i,\textrm{DF}}^{\textrm{EHB}} = 0$ is used for the information decoder and an extra power of $P|h_{si}|^2/|h_{id}|^2$ is exhausted from the battery to ensure that the channel capacity from the source to relay $R_i$ is equal to that from the relay $R_i$ to destination for maximizing the overall channel capacity from the source via relay $R_i$ to destination, namely ${C_{si}} = {C_{id}} = \frac{1}{2}\log ( {1 + \gamma {{\left| {{h_{si}}} \right|}^2}} )$. Therefore, the remaining battery power of relay $R_i$ is obtained as $P_i^s - P|{h_{si}}{|^2}/|{h_{id}}{|^2}$ for ${1 - P _i^s|{h_{id}}{|^2}/(P |{h_{si}}{|^2})} < 0$. Otherwise, when ${1 - P _i^s|{h_{id}}{|^2}/(P |{h_{si}}{|^2})} \ge 0$, it means that all the stored battery power $\mathop P\nolimits_i^s$ would be exhausted for retransmitting the source signal as implied from (56) and (57), and thus the remaining battery power of relay $R_i$ becomes zero for ${1 - P _i^s|{h_{id}}{|^2}/(P |{h_{si}}{|^2})} \ge 0$.}} As a consequence, if the relay $R_i$ is chosen to re-transmit the source signal to the destination, its remaining battery power is given by
\begin{equation}\label{equa59}
\mathop P\nolimits_i^s  = \min\left[\mathop P\nolimits_i^s{\left( {1  - \frac{{P|{h_{si}}{|^2}}}{\mathop P\nolimits_i^s{|{h_{id}}{|^2}}}} \right)^ + },P^{\max}_{b}\right],
\end{equation}
where $P^{\max}_{b}$ represents the power generated from the maximal energy that can be stored in the battery of an EH relay due to the battery capacity limit. Otherwise, all the currently harvested energy $E_i^c$ is accumulated in the EHB of relay $R_i$ and the battery power is given by
\begin{equation}\label{equa60}
\mathop P\nolimits_i^s  = \min\left[\mathop P\nolimits_i^s + \eta P{\left| {{h_{si}}} \right|^2},P^{\max}_{b}\right],
\end{equation}
where $\rho_i = 1$ is used since the relay $R_i$ is not chosen to assist the source-destination transmission and no decoding process is necessary at $R_i$ in this case, leading to that the total received signal is utilized for energy harvesting.}}

\subsection{AF based EHB-OPS-RS Scheme}
In this subsection, we consider the AF protocol for the EH relays and propose an AF based EHB-OPS-RS scheme. In the AF strategy, the relay $R_i$ first divides its received source signal as given by (3) into two fractions $\rho_i$ and $1-\rho_i$, which are used for the energy harvester and information retransmission, respectively. Similarly to (52), the transmit power available at the EH relay $R_i$ is given by
\begin{equation}\label{equa61}
\mathop P\nolimits_i^t  = \mathop \rho \nolimits_i \eta P\mathop {\left| {\mathop h\nolimits_{si} } \right|}\nolimits^2  + \mathop P\nolimits_i^s,
\end{equation}
where $\mathop P\nolimits_i^s$ is the power generated from the energy stored at the battery of $R_i$. Then, the EH relay $R_i$ forwards the remaining fraction $1-\rho_i$ of its received source signal with a scaling factor $G_i$ to the destination at a power of $\mathop P\nolimits_i^t$. Thus, the received signal at the destination is written as
\begin{equation}\label{equa62}
\begin{split}
 \mathop y\nolimits_{id}  &= \mathop h\nolimits_{id} \mathop G\nolimits_i (\mathop {\sqrt {1 - \mathop \rho \nolimits_i } h}\nolimits_{si} \sqrt P \mathop x\nolimits_s  + \mathop n\nolimits_{si} ) + \mathop n\nolimits_{id}  \\
&= \mathop h\nolimits_{id} \mathop G\nolimits_i \mathop {\sqrt {1 - \mathop \rho \nolimits_i } h}\nolimits_{si} \sqrt P \mathop x\nolimits_s  + \mathop h\nolimits_{id} \mathop G\nolimits_i \mathop n\nolimits_{si}  + \mathop n\nolimits_{id},  \\
 \end{split}
\end{equation}
from which the corresponding transmit power is given by
\begin{equation}\label{equa63}
P_i^t = (1 - \mathop \rho \nolimits_i )G_i^2|{h_{si}}{|^2}P + G_i^2{N_0}.
\end{equation}
Using (61) and (63), we can obtain the scaling factor $G_i$ as
\begin{equation}\label{equa64}
{G_i} = \sqrt {\frac{{\mathop \rho \nolimits_i \eta P\mathop {\left| {\mathop h\nolimits_{si} } \right|}\nolimits^2  + P_i^s}}{{(1 - \mathop \rho \nolimits_i )|{h_{si}}{|^2}P + {N_0}}}}  \approx \sqrt {\frac{{\mathop \rho \nolimits_i \eta P\mathop {\left| {\mathop h\nolimits_{si} } \right|}\nolimits^2  + P_i^s}}{{(1 - \mathop \rho \nolimits_i )|{h_{si}}{|^2}P}}},
\end{equation}
where the second approximated equation is obtained in the high SNR region {{such that $(1 - \mathop \rho \nolimits_i )|{h_{si}}{|^2}P \gg N_0$ holds with a high probability}}. Combining (62) and (64) yields the received SNR at the destination as
\begin{equation}\label{equa65}
{\textrm{SNR}}_d = \frac{{G_i^2|{h_{si}}{|^2}|{h_{id}}{|^2}(1 - \mathop \rho \nolimits_i )\gamma }}{{G_i^2|{h_{id}}{|^2} + 1}} = \frac{|{h_{si}}{|^2}|{h_{id}}{|^2}\gamma}{f(\rho_i)},
\end{equation}
where the parameter $f({\rho _i})$ is given by
\begin{equation}\nonumber
f({\rho _i}) = \frac{1}{{G_i^2(1 - {\rho _i})}} + \frac{{|{h_{id}}{|^2}}}{{(1 - {\rho _i})}}.
\end{equation}
{{Substituting $G_i$ from (64) into the preceding equation yields}}
\begin{equation}\label{equa66}
\begin{split}
f({\rho _i}) &=  \frac{1}{{{\rho _i}\eta  + P_i^s/(|{h_{si}}{|^2}P)}} + \frac{{|{h_{id}}{|^2}}}{{(1 - {\rho _i})}}\\
& = \frac{1}{{{\rho _i}\eta  + a}} + \frac{b}{{1 - {\rho _i}}},
\end{split}
\end{equation}
where $a = P_i^s/(|{h_{si}}{|^2}P)$ and $b = |{h_{id}}{|^2}$. Hence, an optimal PSR for the AF based EHB-OPS-RS scheme can be obtained for the sake of maximizing the received SNR of (65), namely
\begin{equation}\label{equa67}
\rho _{i,\textrm{AF}}^{\textrm{EHB}} = \arg \mathop {\max }\limits_{0 \le {\rho _i} \le 1} {\textrm{ SNR}}{_d} = \arg \mathop {\max }\limits_{0 \le {\rho _i} \le 1} {\rm{ }}\frac{|{h_{si}}{|^2}|{h_{id}}{|^2}\gamma}{f(\rho_i)},
\end{equation}
which is equivalent to
\begin{equation}\label{equa68}
\rho _{i,\textrm{AF}}^{\textrm{EHB}} = \arg \mathop {\min }\limits_{0 \le {\rho _i} \le 1} {\rm{ }} f({\rho _i}).
\end{equation}
From (66), the first-order and second-order derivatives of $f({\rho _i})$ with respect to $\rho_i$ are obtained as
\begin{equation}\label{equa69}
\frac{{df({\rho _i})}}{{\partial {\rho _i}}} =  - \frac{\eta }{{{{({\rho _i}\eta  + a)}^2}}} + \frac{b}{{{{(1 - {\rho _i})}^2}}},
\end{equation}
and
\begin{equation}\label{equa70}
\frac{{{d^2}f({\rho _i})}}{{\partial \rho _i^2}} = \frac{{2{\eta ^2}}}{{{{({\rho _i}\eta  + a)}^3}}} + \frac{{2b}}{{{{(1 - {\rho _i})}^3}}}.
\end{equation}
It can be observed from (70) that the second-order derivative is always positive for any ${\rho _i}$ in the range of $0 \le {\rho _i} \le 1$, implying the existence of a unique optimal PSR $\rho _{i,\textrm{AF}}^{\textrm{EHB}}$ to maximize the received SNR at the destination. Moreover, the optimal PSR $\rho _{i,\textrm{AF}}^{\textrm{EHB}}$ should make the first-order derivative become zero, namely
\begin{equation}\label{equa71}
- \frac{\eta }{{{{(\rho _{i,\textrm{AF}}^{\textrm{EHB}}\eta  + a)}^2}}} + \frac{b}{{{{(1 - \rho _{i,\textrm{AF}}^{\textrm{EHB}})}^2}}} = 0,
\end{equation}
from which a closed-form solution to the optimal PSR $\rho _{i,\textrm{AF}}^{\textrm{EHB}}$ is given by
\begin{equation}\label{equa72}
\rho _{i,\textrm{AF}}^{\textrm{EHB}} = {\left[ {\frac{{\sqrt \eta   - a\sqrt b }}{{\sqrt \eta   + \sqrt b \eta }}} \right]^ + },
\end{equation}
where $[x]^+ =\max(x,0)$, $a = P_i^s/(|{h_{si}}{|^2}P)$, and $b = |{h_{id}}{|^2}$. It needs to be pointed out that if no batteries are equipped at the EH relays, namely $P_i^s = 0$, the optimal PSR $\rho _{i,\textrm{AF}}^{\textrm{EHB}}$ of (72) is degraded to ${1}/({{1   + |{h_{id}}| \sqrt \eta }})$. Moreover, in the AF based EHB-OPS-RS scheme, if the EH relay $R_i$ is selected to assist the source-destination transmission, all the stored energy of its battery is used for forwarding the source signal and the remaining battery power at $R_i$ is given by
\begin{equation}\label{equa73}
\mathop P\nolimits_i^s =0.
\end{equation}
Otherwise, all the currently harvested energy is accumulated and the battery power of relay $R_i$ is obtained as
\begin{equation}\label{equa74}
\mathop P\nolimits_i^s  = \min\left[\mathop P\nolimits_i^s + \eta P{\left| {{h_{si}}} \right|^2},P^{\max}_{b}\right],
\end{equation}
which is used for future possible retransmissions of the source signal. Substituting the optimal PSR $\rho _{i,\textrm{AF}}^{\textrm{EHB}}$ from (72) into (65), we can obtain the channel capacity from the source via an EH relay to destination relying on the AF based EHB-OPS-RS scheme as
\begin{equation}\label{equa75}
C_{sd,{\textrm{AF}}}^{{\textrm{EHB}}} = \mathop {\max }\limits_{i = 1,2,...,N} \frac{1}{2}\log\left[1+\frac{|{h_{si}}{|^2}|{h_{id}}{|^2}\gamma}{f(\rho^{\textrm{EHB}}_{i,{\textrm{AF}}})}\right],
\end{equation}
which completes the AF based EHB-OPS-RS scheme.

\section{Numerical Results and Discussions}
In this section, we present numerical outage probability results of the EPS-RS and OPS-RS as well as the AF and DF based EHB-OPS-RS schemes. Following the existing literature [26] and [27], the average channel gains of ${\mathop \sigma \nolimits_{si}^2  = \mathop \sigma \nolimits_{id}^2  = 1}$ and an energy conversion efficiency of $\eta = 0.5$ are assumed in our numerical evaluations. Also, an SNR of $\gamma=15{\textrm{dB}}$, a maximal SNR of $\gamma^{\max}_b = P^{\max}_b/N_0=30{\textrm{dB}}$ generated from the battery of an EH relay due to the battery capacity limit, a data rate of $R=1{\textrm{bit/s/Hz}}$, and the number of EH relays $N=6$ are considered, unless otherwise stated. It is pointed out that the AF based EHB-OPS-RS scheme can be regarded as a variation of the AF based joint power splitting and relay selection framework designed in [40] and [41] originally for two-way relay networks.

\begin{figure}
\centering{}\includegraphics[scale=0.6]{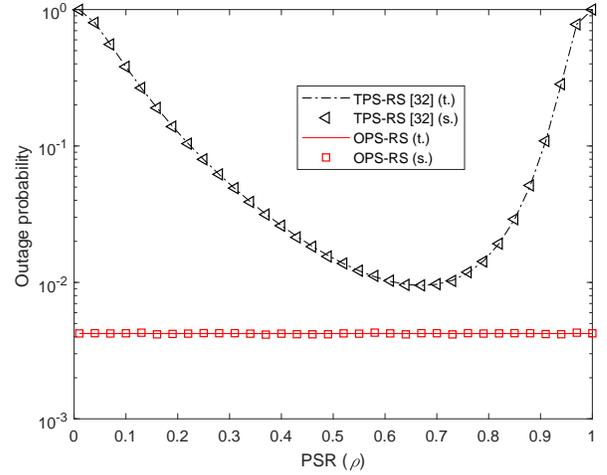}\caption{{{Outage probability versus PSR ${\rho}$ of the traditional power splitting based relay selection (TPS-RS) [32] and proposed OPS-RS schemes with $\gamma=15{\textrm{dB}}$, $R=1{\textrm{bit/s/Hz}}$, $\eta = 0.5$, and $N=6$, where t. and s. represent the theoretical and simulated outage probability results, respectively.}}}
\end{figure}
{{Fig. 3 shows theoretical and simulated outage probabilities versus PSR ${\rho}$ of the traditional power splitting based relay selection (TPS-RS) [32] and proposed OPS-RS schemes with $\gamma=15{\textrm{dB}}$, $R=1$bit/s/Hz, $\eta = 0.5$, and $N=6$. In Fig. 3, the PSR of TPS-RS scheme varies in the interval of $[0,1]$, while an optimized PSR as given by (27) is utilized in the proposed OPS-RS scheme. It is shown from Fig. 3 that with an increasing PSR, the outage probability of TPS-RS decreases first and then increases, demonstrating that a tradeoff exists between the transmission energy and decoding energy. In other words, a minimized outage probability can be achieved for the TPS-RS scheme through an optimization of the PSR. One can also see from Fig. 3 that even the minimized outage probability of TPS-RS with an optimized PSR is much higher than the outage probability of proposed OPS-RS, showing the outage advantage of our scheme. This is because that the PSR optimization of TPS-RS in terms of minimizing the outage probability only utilizes statistical CSI of $h_{si}$ and $h_{id}$, whereas an instantaneous CSI $h_{id}$ is employed in the proposed OPS-RS scheme as implied from (27). In addition, the theoretical and simulated outage probability results of Fig. 3 match well with each other, validating our outage probability analysis.}}

\begin{figure}
\centering{}\includegraphics[scale=0.6]{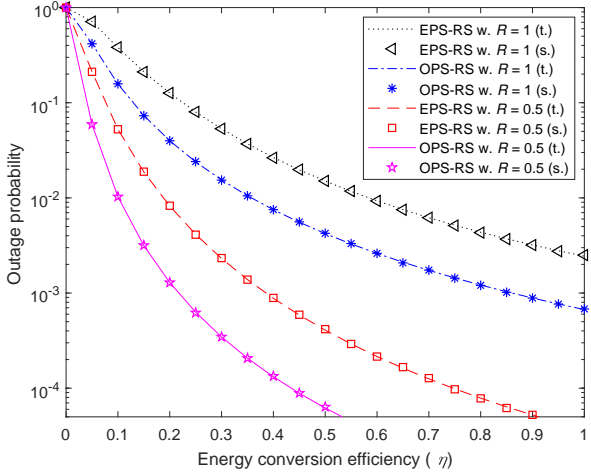}\caption{Outage probability versus energy conversion efficiency ${\eta}$ of the EPS-RS and OPS-RS schemes for different data rates {{with $\gamma=15{\textrm{dB}}$ and $N=6$}}, where t. and s. represent the theoretical and simulated outage probability results, respectively.}
\end{figure}
Fig. 4 shows theoretical and simulated outage probabilities versus energy conversion efficiency ${\eta}$ of the EPS-RS and OPS-RS schemes for different data rates of $R=0.5$bit/s/Hz and $1$bit/s/Hz, where the theoretical outage probabilities of EPS-RS and OPS-RS are obtained by using (23) and (35) and their simulated results are given through Monte-Carlo simulations. It is observed from Fig. 4 that the theoretical outage probability curves match well with the corresponding simulation results, further verifying the correctness of closed-form outage analysis. Fig. 4 also shows that as the energy conversion efficiency ${\eta }$ increases, the outage probabilities of both EPS-RS and OPS-RS decrease accordingly. This is due to the fact that with an increasing $\eta$, more energies are converted from received RF signals for powering the retransmission of the source message to destination, which leads to a lower outage probability.

\begin{figure}
\centering
\includegraphics[scale=0.6]{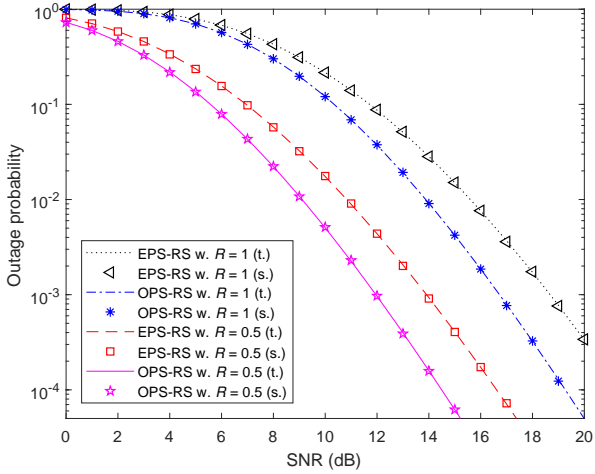}\\
\caption{Outage probability versus SNR of the EPS-RS and OPS-RS schemes for different data rates {{with $\eta = 0.5$ and $N=6$}}, where t. and s. represent the theoretical and simulated outage probability results, respectively.}
\end{figure}

Fig. 5 depicts outage probability versus SNR of the EPS-RS and OPS-RS schemes for different data rates of ${R}=0.5$bit/s/Hz and $1$bit/s/Hz. It can be seen from Fig. 5 that for both cases of ${R}=0.5$bit/s/Hz and ${R}=1$bit/s/Hz, the outage probabilities of EPS-RS and OPS-RS decrease with an increase of SNR, and moreover, the OPS-RS scheme always performs better than the EPS-RS scheme in terms of outage probability. Fig. 5 also demonstrates that the outage probabilities of EPS-RS and OPS-RS increase with an increase of the data rate from ${R}=0.5$bit/s/Hz to $1$bit/s/Hz, implying that the transmission reliability degrades as the data throughput improves, and vice versa.

\begin{figure}
\centering
\includegraphics[scale=0.6]{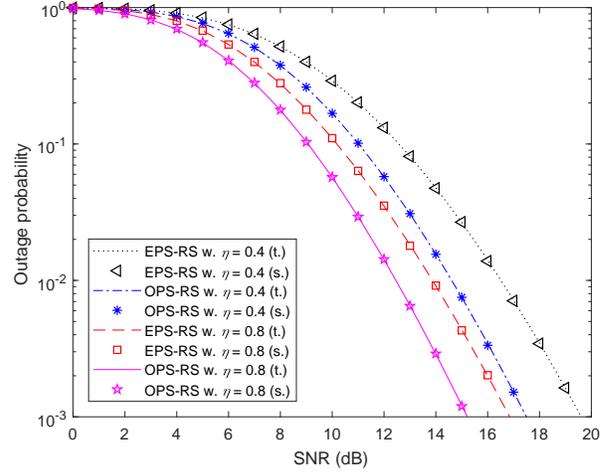}\caption{Outage probability versus SNR of the EPS-RS and OPS-RS schemes for different energy conversion efficiencies with {{with $R=1{\textrm{bit/s/Hz}}$ and $N=6$}}, where t. and s. represent the theoretical and simulated outage probability results, respectively.}
\end{figure}

In Fig. 6, we show outage probability versus SNR of the EPS-RS and OPS-RS schemes for different energy conversion efficiencies of ${\eta}=0.4$ and $0.8$. It is observed from Fig. 6 that given an energy conversion efficiency $\eta$, the outage probabilities of EPS-RS and OPS-RS schemes decrease, as the SNR increases from $\gamma=0$dB to $20$dB. Also, for both cases of ${\eta}=0.4$ and $0.8$, the proposed OPS-RS scheme outperforms traditional EPS-RS method in terms of outage probability across the whole SNR region. Moreover, as the energy conversion efficiency increases from ${\eta}=0.4$ to $0.8$, the outage performance of EPS-RS and OPS-RS improves accordingly.

\begin{figure}
\centering{}\includegraphics[scale=0.6]{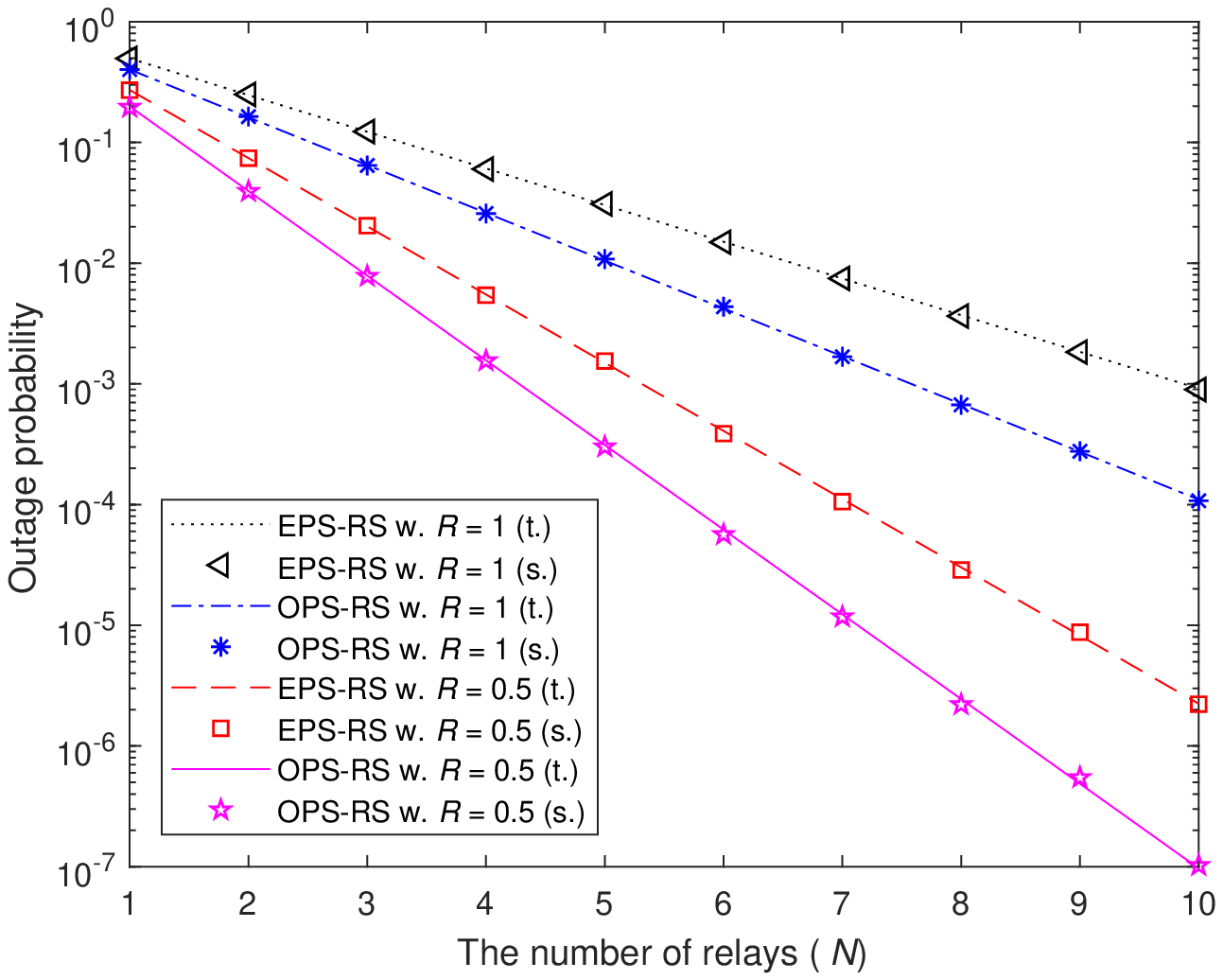}\caption{Outage probability versus SNR of the EPS-RS and OPS-RS schemes for different number of EH relays {{with $\eta = 0.5$ and $R=1{\textrm{bit/s/Hz}}$}}, where t. and s. represent the theoretical and simulated outage probability results, respectively.}
\end{figure}
Fig. 7 depicts outage probability versus SNR of the EPS-RS and OPS-RS schemes for different number of EH relays of $N=4$ and $8$, where both theoretical and simulated outage probabilities are given. It can be seen from Fig. 7 that for both the cases of $N=4$ and $8$, the outage probabilities of EPS-RS and OPS-RS are reduced with an increasing SNR. Moreover, as the number of EH relays increases from $N=4$ to $8$, the outage performance of both schemes improves significantly. One can also observe from Fig. 7 that given an SNR and the number of EH relays $N$, the proposed OPS-RS scheme always outperforms traditional EPS-RS method in terms of the outage probability.

\begin{figure}
\centering{}\includegraphics[scale=0.6]{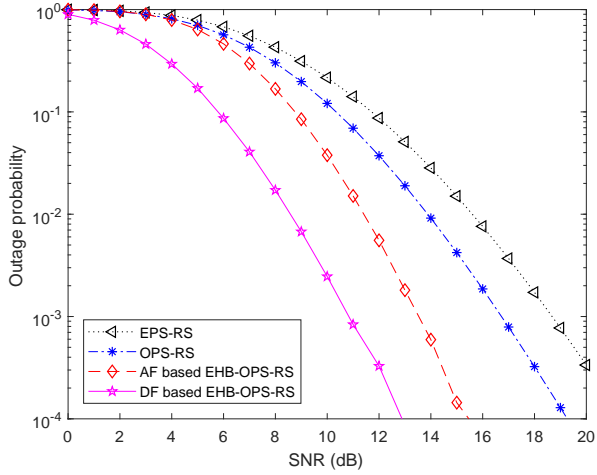}\caption{Outage probability versus the number of EH relays (${N}$) of the EPS-RS and OPS-RS schemes for different data rates {{with $\gamma=15{\textrm{dB}}$ and $\eta = 0.5$}}, where t. and s. represent the theoretical and simulated outage probability results, respectively.}
\end{figure}
Fig. 8 shows outage probability versus the number of EH relays ${N}$ of the EPS-RS and OPS-RS schemes for different data rates of ${R}=0.5$bit/s/Hz and $1$bit/s/Hz. As seen from Fig. 8, for both cases of ${R}=0.5$bit/s/Hz and $1$bit/s/Hz, the outage probability of OPS-RS scheme is always better than that of traditional EPS-RS across the whole region of $N$, and moreover, the performance advantage of proposed OPS-RS over EPS-RS becomes more significant with an increasing number of EH relays. Additionally, as the number of EH relays ${N}$ increases, outage probabilities of both EPS-RS and OPS-RS schemes are reduced substantially, implying significant benefits achieved by the joint power splitting and relay selection in terms of decreasing the outage probability, especially with an increasing number of EH relays.

\begin{figure}
\centering
\includegraphics[scale=0.6]{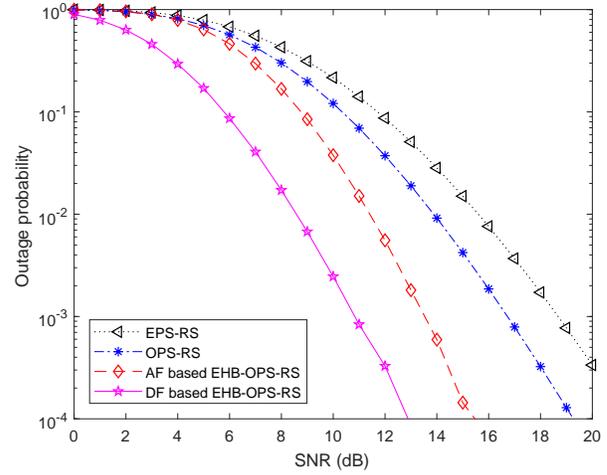}\caption{{{Outage probability versus SNR of the EPS-RS and OPS-RS as well as the AF and DF based EHB-OPS-RS schemes {{with $\eta = 0.5$, $R=1{\textrm{bit/s/Hz}}$, and $N=6$}}.}}}
\end{figure}
In Fig. 9, we provide numerical outage probability comparisons among the EPS-RS and OPS-RS as well as the AF and DF based EHB-OPS-RS schemes. It is pointed out that the outage probability results of AF and DF based EHB-OPS-RS methods are obtained by using (58) and (75) with the aid of Monte-Carlo simulations. As shown in Fig. 9, the AF and DF based EHB-OPS-RS schemes outperform the EPS-RS and OPS-RS methods in terms of their outage probabilities. This is because that in the EPS-RS and OPS-RS schemes, the energies of the relays which have not been selected are just wasted without batteries equipped, which are, however, stored and used for subsequent information transmissions in the AF and DF based EHB-OPS-RS schemes, where the batteries are considered in the EH relays. Fig. 9 also shows that the outage probability of DF-based EHB-OPS-RS is much smaller than that of DF-based  EHB-OPS-RS, implying the outage advantage of the DF protocol over AF strategy.

\begin{figure}
\centering
\includegraphics[scale=0.6]{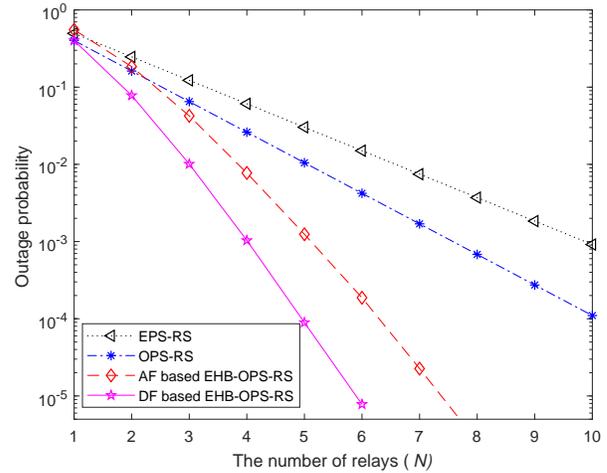}\caption{{{Outage probability versus the number of EH relays (${N}$) of the EPS-RS and OPS-RS as well as the AF and DF based EHB-OPS-RS schemes {{with $\gamma=15{\textrm{dB}}$, $\eta = 0.5$, and $R=1{\textrm{bit/s/Hz}}$}}.}}}
\end{figure}
Fig. 10 depicts outage probability versus the number of EH relays (${N}$) of the EPS-RS and OPS-RS as well as the AF and DF based EHB-OPS-RS schemes. As shown in Fig. 10, the outage probability performance of AF and DF based EHB-OPS-RS schemes are better than that of OPS-RS and EPS-RS methods, further demonstrating the benefit of exploiting the batteries in EH relays for reducing the outage probability of cooperative EH relay communications. One can also observe from Fig. 10 that the DF based EHB-OPS-RS substantially outperforms the AF based EHB-OPS-RS in terms of the outage probability. Moreover, the outage advantage of the DF based EHB-OPS-RS over the AF based EHB-OPS-RS becomes more significant, as the number of EH relays increases.

\section{Conclusion}
In this paper, we investigated the joint power splitting and relay selection for an energy-harvesting (EH) relay network consisting of a source, a destination, and multiple relays that are capable of harvesting energies from their received radio signals. We propose an optimal power splitting based relay selection (OPS-RS) framework, where a closed-form optimal PSR is obtained and an opportunistic relay selection strategy is presented. Also, the traditional equal power splitting based relay selection (EPS-RS) is considered for comparison purposes. We derived closed-form outage probability expressions for both OPS-RS and EPS-RS, based on which their diversity gains are characterized through an asymptotic outage analysis in the high signal-to-noise ratio region. The proposed OPS-RS framework was further extended to an energy-harvesting battery (EHB) enabled cooperative relay scenario, where an amplify-and-forward (AF) based EHB-OPS-RS and a decode-and-forward (DF) based EHB-OPS-RS schemes are proposed for AF and DF relay networks, respectively. Numerical results showed that the outage performance of proposed OPS-RS is better than the traditional EPS-RS, and moreover, the AF and DF based EHB-OPS-RS schemes both perform better than the OPS-RS and EPS-RS methods in terms of the outage probability. Additionally, as the number of EH relays increases, the outage probabilities of AF and DF based EHB-OPS-RS schemes both decrease, but an outage improvement of the DF based EHB-OPS-RS is much more significant than that of the AF based EHB-EPS-RS. In other words, the outage advantage of DF based EHB-OPS-RS over AF based EHB-OPS-RS becomes more substantial with an increasing number of EH relays.

\begin{appendices}
\section{Proof of Proposition 1}
Given an exponential random variable of $x$ with a mean of ${\sigma_{si}^2 }$ and denoting $ \frac{{ \alpha }}{\sigma^2_{id}x} = z$, we can obtain an expected value of ${z}$ as
\begin{equation}\label{equaA.1}
\begin{split}
E(z) &= \frac{\alpha }{{\sigma _{id}^2}}\int_\beta ^\infty  {\frac{1}{{\sigma _{si}^2x}}\exp ( - \frac{x}{{\sigma _{si}^2}})dx}  \\
&= \frac{\alpha }{{\sigma _{si}^2\sigma _{id}^2}}\int_{\frac{\beta }{{\sigma _{si}^2}}}^\infty  {\frac{1}{x}\exp ( - x)dx}  \\
&= \frac{\alpha }{{\sigma _{si}^2\sigma _{id}^2}}Ei(\frac{\beta }{{\sigma _{si}^2}}), \\
\end{split}
\tag{A.1}
\end{equation}
where ${Ei\left( x  \right) = \int_x ^\infty  {\frac{{\mathop e\nolimits^{ - t} }}{t}} dt}$ is known as the exponential integral function. Following (5.1.20) of [47], ${Ei\left(x \right)}$ is bounded to
\begin{equation}\label{equaA.2}
{\frac{1}{2}\exp ( - x )\ln (1 + \frac{2}{x }) \le Ei(x ) \le \exp ( - x )\ln (1 + \frac{1}{x })}, \tag{A.2}
\end{equation}
for ${x  > 0}$. Noting $\beta  = \frac{{2({2^{2R}} - 1)}}{\gamma }$ and letting ${\gamma  \to \infty }$, we can easily obtain
\begin{equation}\label{equaA.3}\tag{A.3}
\mathop {\lim }\limits_{\gamma  \to \infty } \exp ( - \frac{\beta }{{\sigma _{si}^2}}) = 1,
\end{equation}
and
\begin{equation}\label{equaA.4}\tag{A.4}
\mathop {\lim }\limits_{\gamma  \to \infty } \ln (1 + \frac{{2\sigma _{si}^2}}{\beta }) = \ln (\gamma ),
\end{equation}
and
\begin{equation}\label{equaA.5}\tag{A.5}
\mathop {\lim }\limits_{\gamma  \to \infty } \ln (1 + \frac{{\sigma _{si}^2}}{\beta }) = \ln (\gamma ).
\end{equation}
Combining (A.2)-(A.5), we arrive at
\begin{equation}\label{equaA.6}\tag{A.6}
\mathop {\lim }\limits_{\gamma  \to \infty } \frac{1}{2}{\ln (\gamma ) \le \mathop {\lim }\limits_{\gamma  \to \infty } Ei(\frac{\beta }{{\sigma _{si}^2}}) \le \mathop {\lim }\limits_{\gamma  \to \infty } \ln (\gamma )}.
\end{equation}
Substituting (A.6) and $\alpha  = \frac{{2({2^{2R}} - 1)}}{{\eta \gamma }}$ into (A.1), we can obtain
\begin{equation}\label{equaA.7}\tag{A.7}
\mathop {\lim }\limits_{\gamma  \to \infty }\frac{{({2^{2R}} - 1)\ln (\gamma )}}{{\sigma _{si}^2\sigma _{id}^2\eta \gamma }} \le \mathop {\lim }\limits_{\gamma  \to \infty }  E(z)
\le \mathop {\lim }\limits_{\gamma  \to \infty }\frac{{2({2^{2R}} - 1)\ln (\gamma )}}{{\sigma _{si}^2\sigma _{id}^2\eta \gamma }},
\end{equation}
from which one can readily have
\begin{equation}\label{equaA.8}\tag{A.8}
0 \le \mathop {\lim }\limits_{\gamma  \to \infty }  E(z)
\le 0,
\end{equation}
which leads to
\begin{equation}\label{equaA.9}\tag{A.9}
\mathop {\lim }\limits_{\gamma  \to \infty } E(z) = 0.
\end{equation}
It can be seen from (A.9) that the expected value of $z$, ${E(z)}$, converges to zero for ${\gamma  \to \infty }$.

Moreover, the expected value of $z^2$ can be obtained as
\begin{equation}\label{equaA.10}\tag{A.10}
\begin{split}
E({z^2}) &= \frac{{{\alpha ^2}}}{{\sigma _{id}^4}}\int_\beta ^\infty  {\frac{1}{{\sigma _{si}^2{x^2}}}\exp ( - \frac{x}{{\sigma _{si}^2}})dx}  \\
&= \frac{{{\alpha ^2}}}{{\sigma _{si}^2\sigma _{id}^4}}\int_{{\beta }}^\infty  {\exp ( - \frac{x}{\sigma^2_{si}})d( - \frac{1}{x})}  \\
&= \frac{{{\alpha ^2}}}{{\sigma _{si}^2\sigma _{id}^4}}\left( {\frac{{1}}{\beta }\exp ( - \frac{\beta }{{\sigma _{si}^2}}) - \frac{1}{{\sigma _{si}^2}}Ei(\frac{\beta }{{\sigma _{si}^2}})} \right). \\
\end{split}
\end{equation}
Substituting (A.3) and (A.6) into (10) yields
\begin{equation}\label{equaA.11}\tag{A.11}
\begin{split}
&\mathop {\lim }\limits_{\gamma  \to \infty } \frac{{{\alpha ^2}}}{{\beta \sigma^2_{si}\sigma _{id}^4}} - \frac{{{\alpha ^2}\ln (\gamma )}}{{\sigma _{si}^4\sigma _{id}^4}} \le \mathop {\lim }\limits_{\gamma  \to \infty }E({z^2})\\
&\le \mathop {\lim }\limits_{\gamma  \to \infty } \frac{{{\alpha ^2}}}{{\beta \sigma^2_{si}\sigma _{id}^4}} - \frac{{{\alpha ^2}\ln (\gamma )}}{{2\sigma _{si}^4\sigma _{id}^4}},
\end{split}
\end{equation}
for ${\gamma  \to \infty }$. Substituting $\alpha  = \frac{{2({2^{2R}} - 1)}}{{\eta \gamma }}$ and $\beta  = \frac{{2({2^{2R}} - 1)}}{\gamma }$ into (A.11) and ignoring higher-order infinitesimals give
\begin{equation}\label{equaA.12}\tag{A.12}
\mathop {\lim }\limits_{\gamma  \to \infty } E({z^2}) = \frac{{2({2^{2R}} - 1)}}{{\sigma^2_{si}\sigma _{id}^4{\eta ^2}}} \cdot \frac{1}{\gamma } ,
\end{equation}
which leads to
\begin{equation}\label{equaA.13}\tag{A.13}
\mathop {\lim }\limits_{\gamma  \to \infty } E({z^2}) = 0.
\end{equation}
Combining (A.9) and (A.13), the variance of ${z}$ can be obtained as
\begin{equation}\label{equaA.14}\tag{A.14}
Var(z) = E(\mathop z\nolimits^2 ) - \mathop {\left[ {E(z)} \right]}\nolimits^2  = 0,
\end{equation}
for ${\gamma  \to \infty }$.

As shown from (A.9) and (A.14), as the SNR $\gamma$ goes to infinity, both the mean and variance of the random variable ${z}$ approach to zero. This means that the random variable $z$ approaches to zero for ${\gamma  \to \infty }$. Therefore, noting $ \frac{{ \alpha }}{\sigma^2_{id}x} = z$ and considering ${\gamma  \to \infty }$, we can obtain
\begin{equation}\label{equaA.15}\tag{A.15}
\frac{{ \alpha }}{\sigma^2_{id}x} = 0,
\end{equation}
which completes the proof of Proposition 1.

\end{appendices}

\begin{IEEEbiography}[{\includegraphics[width=1in,height=1.25in]{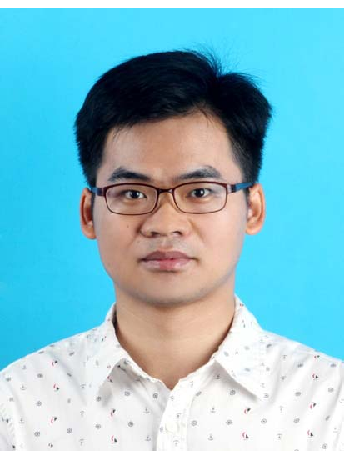}}]{Yulong Zou} (SM'13) is a Full Professor and Doctoral Supervisor at the Nanjing University of Posts and Telecommunications (NUPT), Nanjing, China. He received the B.Eng. degree in information engineering from NUPT, Nanjing, China, in July 2006, the first Ph.D. degree in electrical engineering from the Stevens Institute of Technology, New Jersey, USA, in May 2012, and the second Ph.D. degree in signal and information processing from NUPT, Nanjing, China, in July 2012.

Dr. Zou was awarded the 9th IEEE Communications Society Asia-Pacific Best Young Researcher in 2014. He has served as an editor for the IEEE Communications Surveys \& Tutorials, IEEE Communications Letters, EURASIP Journal on Advances in Signal Processing, IET Communications, and China Communications. In addition, he has acted as TPC members for various IEEE sponsored conferences, e.g., IEEE ICC/GLOBECOM/WCNC/VTC/ICCC, etc.
\end{IEEEbiography}

\begin{IEEEbiography}[{\includegraphics[width=1in,height=1.25in]{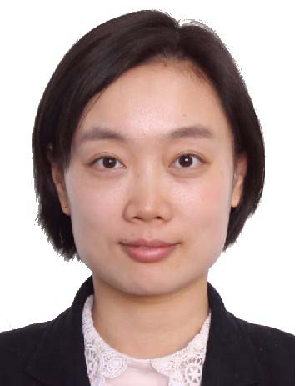}}]{Jia Zhu} is a Full Professor at the Nanjing University of Posts and Telecommunications (NUPT), Nanjing, China. She received the B.Eng. degree in Computer Science and Technology from the Hohai University, Nanjing, China, in July 2005, and the Ph.D. degree in Signal and Information Processing from the Nanjing University of Posts and Telecommunications, Nanjing, China, in April 2010. From June 2010 to June 2012, she was a Postdoctoral Research Fellow at the Stevens Institute of Technology, New Jersey, the United States. Her general research interests include the cognitive radio, physical-layer security, and communications theory.\end{IEEEbiography}

\begin{IEEEbiography}[{\includegraphics[width=1in,height=1.25in]{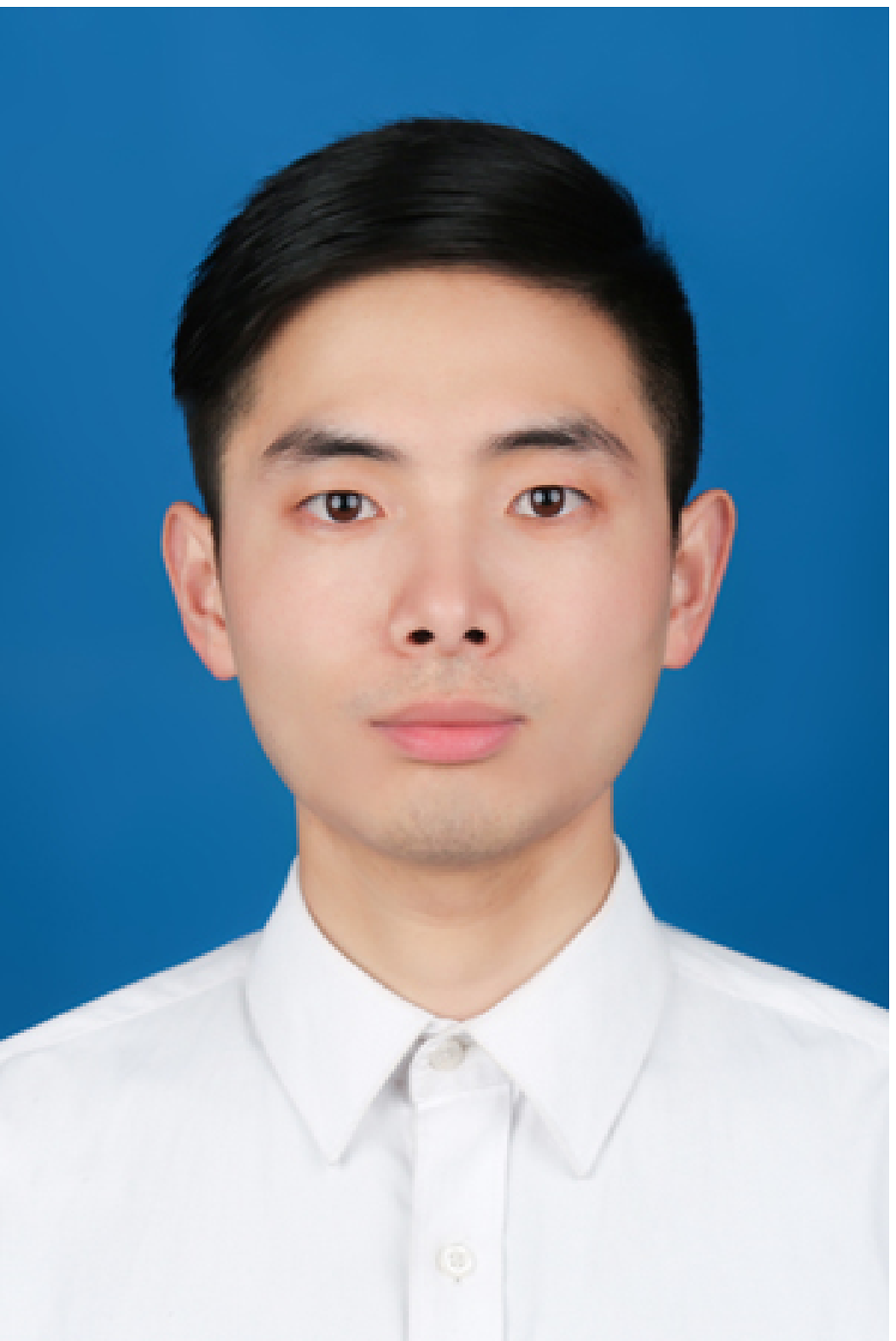}}]{Xiao Jiang} received the M. Eng degree in signal and information processing from the Nanjing University of Posts and Telecommunications (NUPT), Nanjing, China, in April 2019. His research interests include energy harvesting, physical-layer security, and cooperative communications.\end{IEEEbiography}

\end{document}